\documentclass[twocolumn,english,prl]{revtex4-1}
\usepackage{color}
\usepackage{graphicx}
\usepackage[T1]{fontenc}
\usepackage{float}
\usepackage{textcomp}
\usepackage{amsmath}
\usepackage{amssymb}
\usepackage{graphicx}
\usepackage{esint}
\usepackage{natbib}
\usepackage{color}

\begin{document}


\title{Electrostatic quantum dot confinement in phosphorene}


\author{Bart\l{}omiej Szafran}

\affiliation{AGH University of Science and Technology, Faculty of Physics and
Applied Computer Science,\\
 al. Mickiewicza 30, 30-059 Krak\'ow, Poland}

\begin{abstract}
We consider states localized by electrostatic potentials in phosphorene
using atomistic tight-binding approach. From the tight-binding spectra of states confined in parabolic potential we extract effective masses for conduction band electrons moving along the armchair and zigzag crystal directions
by a fit to the harmonic oscillator spectrum.
The masses derived in this way are used for a simple single-band effective-mass model which, as we find, reproduces very well the tight-binding energy spectra in external magnetic field, the confined probability densities,  and the interaction effects. We study the confined states in the conduction band and find that both methods already for small quantum dots produce Wigner crystallization with  separated  single-electron islands. The effective-mass model works with a slightly worse precision for confined states of the valence band. The continuum approach deviates considerably from the tight-binding model only in the limit of extremely strong lateral confinement. 
\end{abstract}

\maketitle

\section{introduction}

Black phosphorus (BP) is a layered semiconductor material that attracts a growing attention for its basic properties and possible applications \cite{aniso,14,otpo,rev,fukuoka15}. BP with direct band gap  \cite{anisobp}  that  falls within the visible range for few-layer systems  \cite{prbgap,thbgnl,fofo}  
is investigated for optoelectronic applications   \cite{otpo}. Separation of few-layer material down to a monolayer  (phoshorene) \cite{fosf14}  is now routinely accomplished. The lateral confinement for optics 
is achieved in nanocrystal \cite{Lee17,Sun15,Wang18,BPQD18,Saroka,s2,s3,pet1} quantum dots (QDs) that support confinement of both electrons and holes with by a mere finite size of the medium.

 Gating of BP  grown on nonmetallic substrates allows for fabrication of field effect transistors \cite{Bfet14} that recently reached a long-term air stability  at room temperature \cite{he19,bl19}. The integer \cite{hhe,ehe} and fractional quantum Hall \cite{qhe} effects,
the latter being a fingerprint of strong correlations between interacting carriers, has already been reported  \cite{qhe}.   The electrostatic QDs in bulk semiconductors \cite{eqd}, bilayer graphene \cite{prx}
or carbon nanotubes \cite{laird} with their clean  confinement independent  of the nanocrystal edges, allow  for precise studies of localized states,  energy spectra, coherence times \cite{eqd}, electron-electron interactions  \cite{reim} as well as for control of the charge \cite{ceqd1,ceqd2} and spin \cite{eqds} degrees of freedom. 
Although the advanced stage of BP field effect transistors has been reached  \cite{ehe,Bfet14,he19,bl19,hhe,qhe} 
 so far there are neither experimental nor theoretical literature on electrostatic QD confinement  in BP. 

In this paper we study a single electron and an electron pair  confined in a phosphorene QD by an external potential that can be induced by electrostatic confinement.
The electrostatic confinement potential is usually parabolic near its minimum \cite{lis}.
For materials with a parabolic dispersion relation the quantum harmonic oscillator is formed in this way,
with the Fock-Darwin dependence on the external magnetic field for the isotropic case  \cite{reim}.  The anisotropy of phoshorene crystal structure  \cite{rev} is translated to the in-plane anisotropy of the valence and conduction bands \cite{aniso,anisobp,18,tibikast1,tibikast2}, with the effective masses along the zigzag direction that are much larger than in the armchair direction \cite{18}. 
The electrostatic BP quantum dots are promising for observation of strong interaction effects due to large
effective masses in one of the directions and for tuning the interaction by orientation of the 
confinement potential anisotropy, and not the size of the dot, as for materials with isotropic effective masses. 
The BP energy bands deviate from parabolic \cite{prbli19,kdotp2,kdotp1} near the conduction and valence band extrema.
Due to the non-parabolicity  a precise continuum description  calls for $k \cdot p$ modelling of the low energy bands with the coupling between the conduction and valence bands  \cite{kdotp2,kdotp3,kdotp4,kdotp5}. The coupling is relatively weak for monolayer BP due to the large band gap \cite{prbgap,fofo}. 
In phosphorene the Landau levels are nearly linear on the external magnetic field and the non-linear corrections turn out to be small \cite{kdotp6,small}. On the other hand already for bilayer phosphorene the 
energy spectra in external magnetic field are very complex, non-linear and corresponding to different effective masses for  each level \cite{kdotp6}.  The relatively simple form of Landau levels for phosphorene  \cite{kdotp6,small} motivated us to look for description of the confined states in a parabolic single-band effective mass modeling.
We extract the effective masses from the tight-binding model by imposing a parabolic confinement potential. Next we use the effective masses as external potential parameters
for which the tight-binding calculations approximate best the quantum harmonic osillator spectrum for $B=0$ with its characteristic degeneracies and energy spacings. 
  Although the tight-binding spectrum deviates from the exact quantum harmonic oscillator, the single and two-electron levels calculated by the tight-binding method  can be surprisingly well reproduced by the simple single-band effective-mass model.

\section{Theory}

\subsection{single-electron Hamiltonian}

\begin{figure}
 \includegraphics[width=1\columnwidth]{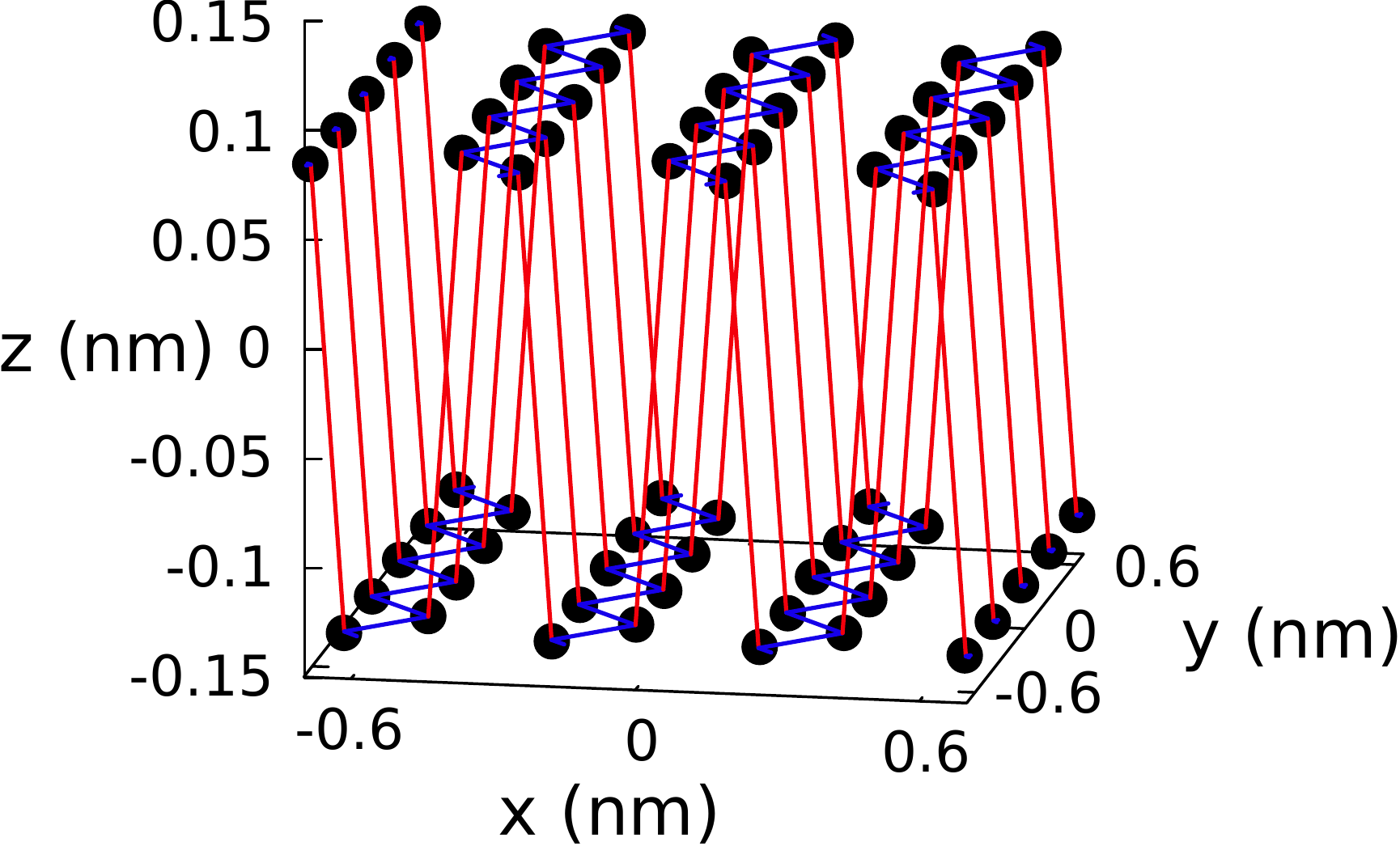} 
\caption{Crystal structure of phosphorene with zigzag lines extended along the $y$ direction. 
The zigzag bonds appear on two parallel planes shifted in the $z$ direction. 
The links between the atoms show the pairs of the largest absolute values of the hopping energy (see Table I).
 The nearest neighbors within the zigzag chain are linked with the blue lines that are spaced by 2.22 \AA\; with the hopping energy of $-1.22$ eV. The red lines show the nearest-neighbor links between the zigzag chains
of separate planes that are 2.24 \AA\; long with the hopping energy of $3.665$ eV.
}
 \label{crystal}
\end{figure}

We consider a monolayer BP (see Fig. \ref{crystal}) with zigzag lines extended along the $y$ direction 
and the armchair lines along the $x$ axis. We use the effective tight-binding Hamiltonian of Ref. \cite{tibikast1}, 
\begin{eqnarray}
H_{TB}=\sum_{kl } t_{kl} p_{kl} c_{k}^\dagger c_{l} 
 +\sum_{k} V_k c^\dagger_{k}c_{k}+\frac{g\mu_B B}{2}\sigma_z, \label{hb0}
\end{eqnarray}
where the first sum describes the hopping between the neighbor atoms,
and $p_{kl}$ is the Peierls phase that introduces the orbital effects of the magnetic field
$p_{kl}=e^{i\frac{e}{\hbar}\int_{\vec{r_k}}^{\vec{r_l}}\vec {A}\cdot \vec {dl}}$ to the hopping elements.
We use the  symmetric gauge ${\bf A}=(A_x,A_y,A_z)=(-By/2,Bx/2,0)$ for the perpendicular magnetic field $(0,0,B)$
The hopping integrals $t_{kl}$ adopted from Ref. \cite{tibikast1} are listed in Table \ref{tab1}.
The pairs of ions that correspond to the two largest absolute values of $t_{kl}$ are linked in Fig. \ref{crystal} by blue ($t_{kl}=-1.22$ eV) and red ($t_{kl}=3.665$ eV) lines. 
The second sum of Eq. (\ref{hb0}) introduces the external potential, and the last term stands for the Zeeman effect with the Land\'e factor $g=2$, and Bohr magneton $\mu_B$.
We consider   a finite rectangular flake of phosphorene 
with a side length of 87 nm in the $x$ direction and 44 nm in the $y$ direction
with over 100 thousand ions. The size of the flake is larger than the confinement area of low-energy states considered below.

For the external potential we use the harmonic oscillator potential 
\begin{equation} V(x,y)=\frac{1}{2} m_x\omega^2 x^2+\frac{1}{2}m_y\omega^2 y^2, \label{pote} \end{equation}
where $\hbar \omega$ is the oscillator energy,
the $m_x$ and $m_y$ are fit parameters for the electron effective masses in the armchair and zigzag directions, respectively.
For $m_x\neq m_y$ potential (\ref{pote}) is anisotropic.
In Eq. (\ref{hb0}) $V_k$ is the potential on $k$th ion, $V_k=V(x_k,y_k)$, where $x_k$, and $y_k$ are the coordinates of the $k$th ion. 
We establish $m_x$ and $m_y$ in the potential of Eq. (\ref{pote}) as fit parameters
for the tight-binding spectrum to reproduce the 
 harmonic oscillator spectrum at $B=0$.

\subsection{confined electron pair in the tight-binding approach}
We calculate the spectrum for a confined electron pair using 
 the Hamiltonian,
\begin{equation}
  {H}_{2e}=\sum_{i}{d}^{\dagger}_{i}{d}_{i}E_i +\frac{1}{2}\sum_{ijkl}{d}^{\dagger}_{i}{d}^{\dagger}_{j}{d}_{k}{d}_{l}V_{ijkl},
  \label{Htwo}
\end{equation}
where ${d}^{\dagger}_{i}$ is the electron creation operator for the single-electron energy level $E_i$  and the Coulomb matrix elements read
\begin{equation}
V_{ijkl}=\kappa\langle{\psi_{i}(\mathbf{r_1})\psi_{j}(\mathbf{r_2})\frac{1}{|\mathbf{r_{12}}|}}{\psi_{k}(\mathbf{r_1})\psi_{l}(\mathbf{r_2})}\rangle, 
\label{coulF}
\end{equation}
where $\kappa=e^2/(4\pi\epsilon\epsilon_0)$, and $\psi$'s standing for eigenstates of single-electron Hamiltonian (1). We consider that the phoshorene
is embedded in Al$_2$O$_3$ matrix and use its dielectric constant  $\epsilon_0=9.1$.
We integrate the Coulomb elements for the single-electron wave functions $\psi$ spanned by the
atomic orbitals $3p_z$ of $P$ atoms,
\begin{equation}
\psi_i({\bf r}_i)=\sum_{k} C^i_{k} p_z^k({\bf r}_1),
\end{equation}
\begin{eqnarray}
V_{ijkl}&=&\kappa\langle\psi_{i}({\bf r_{1}})\psi_{j}({\bf r_{2}})|\frac{1}{|{r_{12}}|}|\psi_{k}({\bf r_{1}})\psi_{l}({\bf r_{2}})\rangle\nonumber \\
&=&\kappa\sum_{\substack{a,\sigma_{a};b,\sigma_{b};\\ c,\sigma_{c};d,\sigma_{d} }}C_{a,\sigma_{a}}^{i*}C_{b,\sigma_{b}}^{j*}C_{c,\sigma_{c}}^{k}C_{d,\sigma_{d}}^{l}\delta_{\sigma_{a};\sigma_{d}}\delta_{\sigma_{b};\sigma_{c}}\times \nonumber\\&&\langle p_{z}^{a}({\bf r}_{1})p_{z}^{b}({\bf r_{2}})|\frac{1}{|{r_{12}}|}|p_{z}^{c}({\bf r_{1}})p_{z}^{d}({\bf r_{2}})\rangle.
\end{eqnarray}
For the Coulomb integral we apply the two-center approximation \cite{c2}
$ \langle p_{z}^{a}({\bf r}_{1})p_{z}^{b}({\bf r_{2}})|\frac{1}{|{r_{12}}|}|p_{z}^{c}({\bf r_{1}})p_{z}^{d}({\bf r_{2}})\rangle=\frac{1}{r_{ab}} \delta_{ac}\delta_{bd}$ for $a\neq b$. 
The  on-site integral ($a=b$) is calculated with $3p_z$  atomic orbitals,  $p_z({\bf r})=N z \left(1-\frac{Z r}{6}\right)\exp(-Zr/3)$, where $N$ is the normalization constant and $Z$ is the effective screened P nucleus charge as seen by $3p_z$ electrons. The single-center integral can then be 
calculated analytically, $I_{a=b}=\frac{3577}{46080}Z$ in atomic units. The Slater screening rules for $3p$ electrons in P atoms produce $Z=4.8$, then $I_{a=b}=10.14$ eV.

The Hamiltonian (\ref{Htwo}) is diagonalized with the configuration interaction approach in the basis of up to $\sim 1000$ two-electron Slater determinants
constructed from the lowest-energy conduction-band eigenfunctions
of the single-electron Hamiltonian (\ref{hb0}).

\begin{table}
\begin{tabular}{l|llllll} 
 $r_{kl}($\AA$)$& $2.22$& $2.24$ &$3.34$ & $3.47$ & $4.23$ \\ 
 $t_{kl}($eV$)$& -1.22& 3.665&-0.205& -0.105 & -0.055 
\end{tabular}
\caption{Hopping parameters for the effective tight-binding model of Ref. \cite{tibikast1}. The value
of the hopping parameter $t_{ij}$ is given below the distance between P ions $r_{ij}$.}
\label{tab1}
\end{table}

\subsection{single-band effective-mass Hamiltonian}
For description of the system in the effective-mass Hamiltonian we take the single-band
approximation with the  Hamiltonian,
\begin{eqnarray}
H_{em}&=&\frac{\left(-i\hbar \frac{\partial }{\partial x}+e{A_x}\right)^2}{2m_x}+\frac{\left(-i\hbar \frac{\partial }{\partial y}+e{A_y}\right)^2}{2m_y}\nonumber \\ &+&V(x,y)+\frac{g\mu_B B}{2}\sigma_z. \label{1eh}
\end{eqnarray}
To diagonalize this Hamiltonian we employ the finite-difference method with 
gauge-invariant discretization of Ref. \cite{governale} and Peierls phases that account for the orbital effects of the magnetic field.
For the mesh spacing $\Delta x$ in both $x$ and $y$ directions finite difference Hamiltonian 
defined by its action on the wave function $\Psi_{\mu,\nu}=\Psi(x_\mu, x_\nu)=\Psi(\mu \Delta x, \nu \Delta x),$
\begin{eqnarray}
H_{fd} \Psi_{\mu,\nu}&\equiv& \frac{\hbar^2}{2m_x \Delta x^2}\left(2\Psi_{\mu,\nu}-C_y \Psi_{\mu,\nu-1}-C_y^*\psi_{\mu,\nu+1}\right)\nonumber \\
&+&\frac{\hbar^2}{2m_y \Delta x^2}\left(2\Psi_{\mu,\nu}-C_x \Psi_{\mu-1,\nu}-C_x^*\psi_{\mu+1,\nu}\right)\nonumber \\ &+& V_{\mu,\nu}\Psi_{\mu,\nu}, \label{numei}
\end{eqnarray}
with $C_x=\exp(-i\frac{e}{\hbar}\Delta x A_x)$ and $C_y=\exp(-i\frac{e}{\hbar}\Delta x A_y)$. 

With the eigenstates of Hamiltonian  (\ref{numei})
we calculated the two-electron spectrum in the manner discussed above for the
tight-binding method. The only difference is the on-site integral which is evaluated numerically
\begin{eqnarray} I_{a=b}&=&\int_{-\Delta x/2}^{\Delta x/2} dx_1\int_{-\Delta x/2}^{\Delta x/2} dy_1\int_{-\Delta x/2}^{\Delta x/2} dx_2\int_{-\Delta x/2}^{\Delta x/2} dy_2\nonumber \\&&\frac{1}{\sqrt{(x_1-x_2)^2+(y_1-y_2)^2}}\end{eqnarray} with the Monte Carlo method. With the finite difference mesh we cover the same area as in the 
tight-binding approach and use $\Delta x=0.2$ nm. 

\section{Results}
\subsection{confined electrons of the conduction band}
\subsubsection{TB spectrum fit for the effective masses}
We determine the effective masses by a fit of the tight-binding
spectrum obtained for Hamiltonian (\ref{hb0}) to the harmonic oscillator spectrum with potential (\ref{pote}).
Note, that the fit involves variation of the confinement potential of Eq.(\ref{pote}) as the masses are varied.
At $B=0$ the harmonic oscilator  the $n$th excited energy level  is $n+1$ fold degenerate (spin excluded) with the energy $n\hbar \omega$ above the ground state. 
For the fit we  took $\hbar\omega=10$ meV and considered 15 lowest energy levels obtained with the tight-binding approach, including the ground-state. 
We looked for a minimal sum of squares of the energy difference between electron energy levels obtained with the tight-binding approach and the corresponding quantum harmonic oscillator energy levels.   The best fit was obtained for $m_x=0.17037m_0$ and $m_y=0.85327 m_0$,
with the conduction band effective masses that fall within the range of the values indicated in the literature, e.g.,
$m_x/m_0=0.148$ \cite{nano},  0.16 \cite{016}, 0.166 \cite{kdotp6}, 0.17 \cite{18}, 0.2 \cite{scre2}, 
and $m_y/m_0=0.846$ \cite{kdotp6}, 1.12 \cite{18}, 1.237 \cite{nano}, 1.24 \cite{016}. In the literature also values as large as $m_y=6.89m_0$ are used \cite{scre2}. Values large as the latter
are obtained for strained phosphorene layers \cite{ntstrain}.
The optimal spacings between the excited states and the ground states are given in Table \ref{tab2}. 
The degeneracy of the excited energy levels is only approximate for the best fit. 
In Table \ref{tab2} the largest energy diffrence $|\Delta E|$ between the tight-binding harmonic oscillator levels are 0.033 meV, 0.034 meV, 0.041 meV, and 0.117 meV  for $n=1,2,3$, and $4$, respectively.

\begin{table}
\begin{tabular}{l|lllll|l} 
$n\hbar \omega$ & $\Delta E_{TB}$ & &&&&$\sqrt{\sum_i (E^{ex}_i-E^{tb}_i)^2}$ \\
\hline
 10 & 10.000 & 10.033 &&&&0.0330 \\  20 & 20.000 & 20.001 & 20.034&&&0.0340 \\ 30&  29.959 & 29.961 & 30.013 & 30.021 && 0.0617 \\  40 & 39.883 &39.893 & 39.980 &39.996 & 40.022 & 0.1614 \\
\end{tabular}
\caption{The energy spacings from the ground state in meV for  14 excited states of the conduction band with the tight-binding method for confinement potential given by Eq. (\ref{pote}). The electron effective masses fitted to the harmonic oscillator spectrum $m_x=0.17037m_0$, $m_y=0.85327m_0$ for $\hbar\omega=10$ meV and $B=0$.
In the subsequent rows we group the nearly degenerate energy levels that for the quantum harmonic oscillator 
should be placed exactly at the energy of $n\hbar\omega$ above the ground state for $n=1,2,3$ and 4.
The last column shows the root of the square distance between the harmonic oscillator spectrum and the tight-binding
result for given $n$ in milielectronovolts.
} \label{tab2}
\end{table}

 \begin{figure}
\begin{tabular}{ll}
 \includegraphics[width=.51\columnwidth]{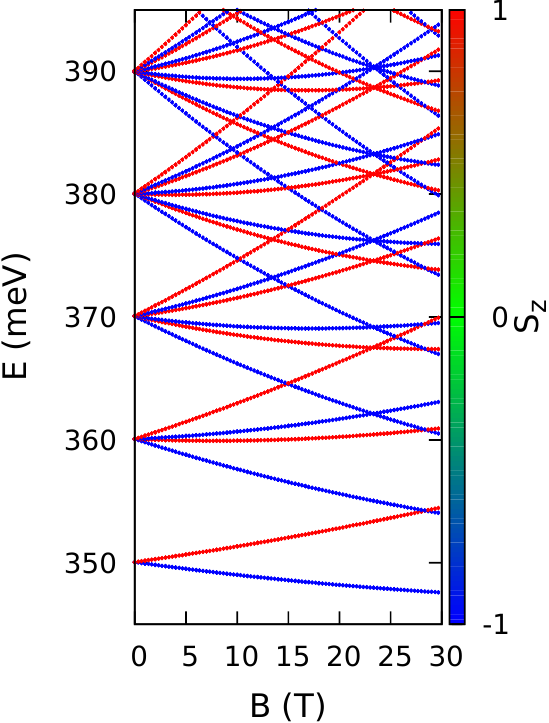} \put(-93,50){(a)}  &
 \includegraphics[width=.5\columnwidth]{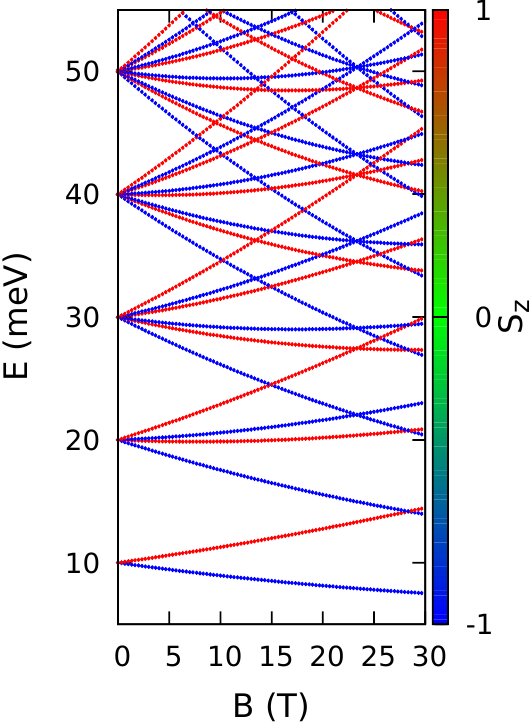}\put(-93,50){(c)}\\
 \includegraphics[width=.51\columnwidth]{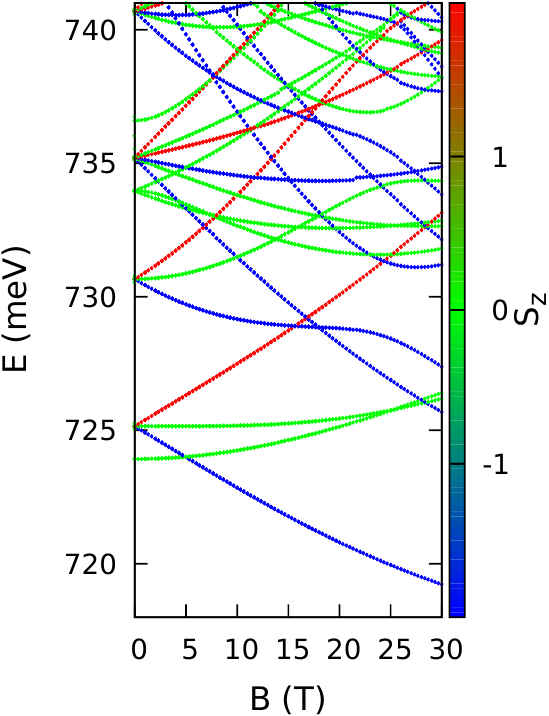}\put(-93,50){(b)} &
 \includegraphics[width=.51\columnwidth]{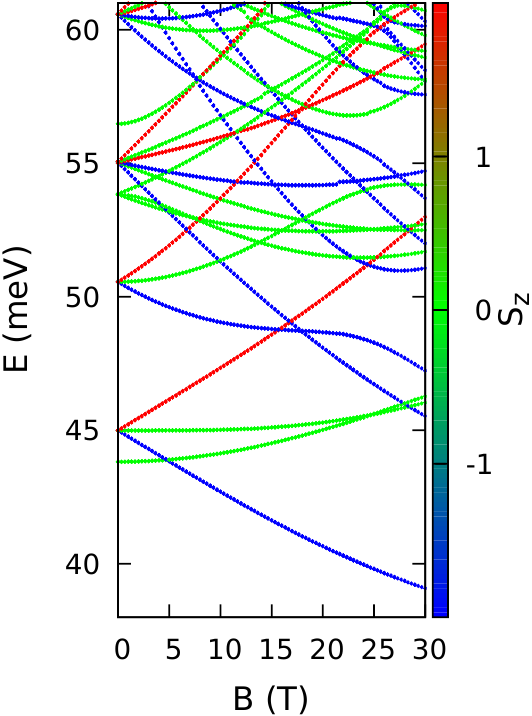}\put(-93,50){(d)}  

\end{tabular}

\caption{Tight-binding (a,b) spectrum and the results of the single-band
effective-mass approximation (c,d) for a single (a,c) and two electrons (b,d)
confined in potential (\ref{pote}) with $m_x=0.17037m_0$, $m_y=0.85327m_0$ and $\hbar\omega=10$ meV
in the external magnetic field. The color of the line indicates the $z-$ component of the spin.}
 \label{jedenidwae}
\end{figure}

 \begin{figure}
\begin{tabular}{ll}
 \includegraphics[width=.5\columnwidth]{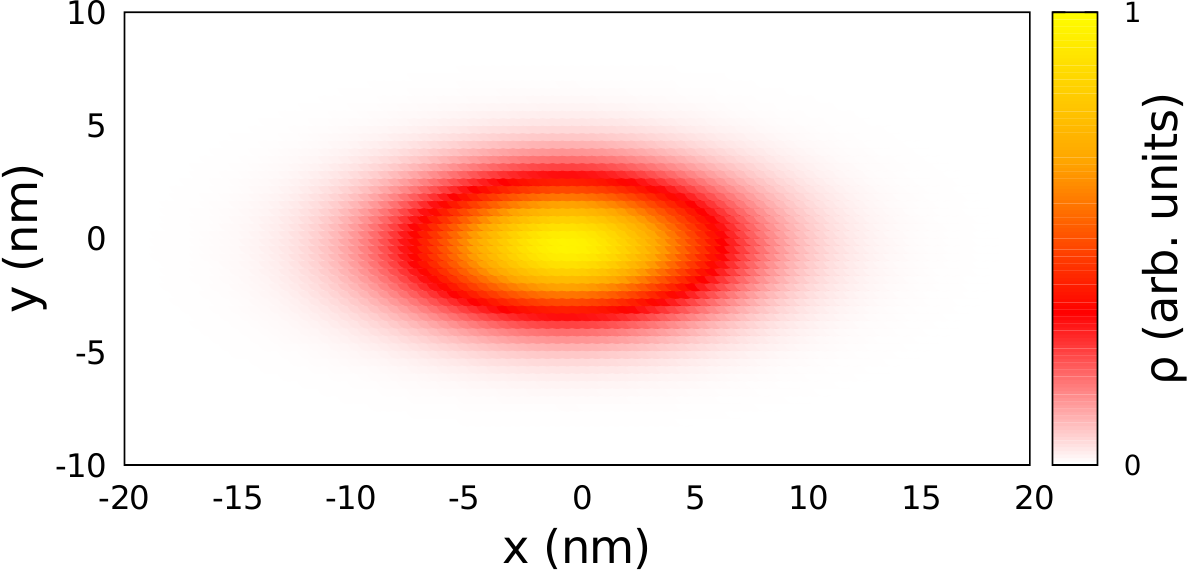}\put(-105,42){(a)} &
 \includegraphics[width=.5\columnwidth]{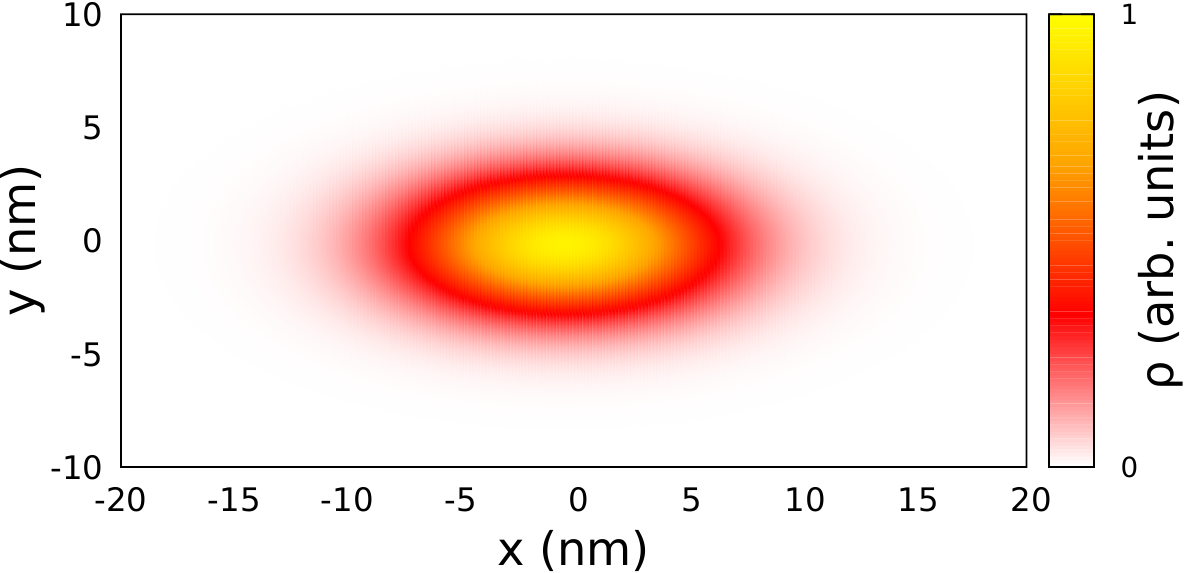}\put(-105,42){(b)} 
\\
 \includegraphics[width=.5\columnwidth]{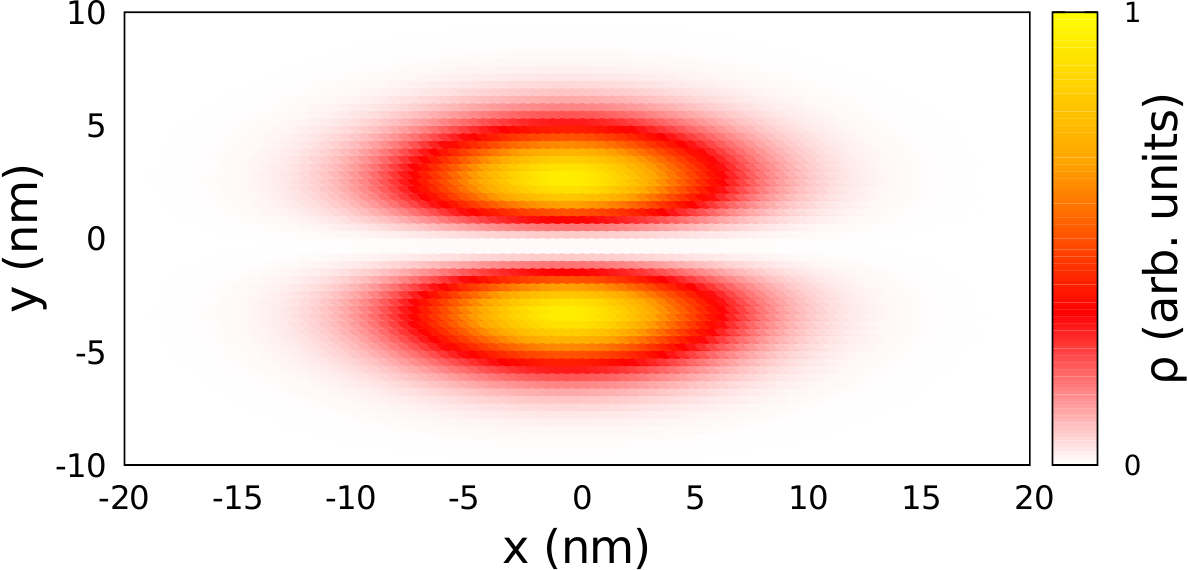} \put(-102,41){(c)}&  
 \includegraphics[width=.5\columnwidth]{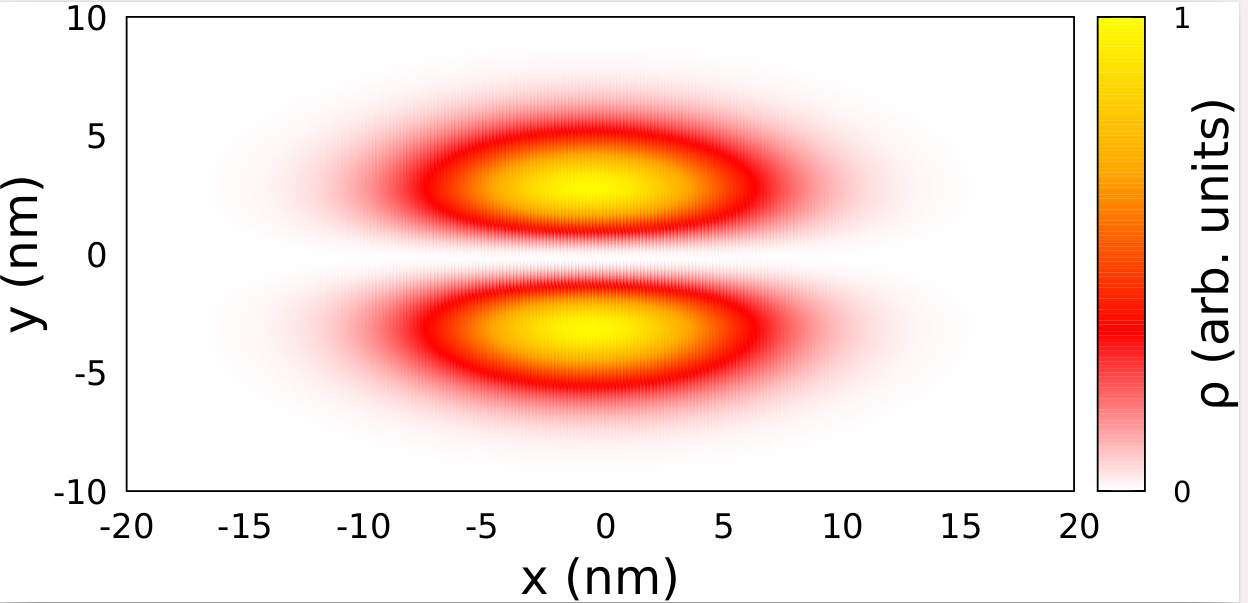}\put(-102,41){(d)} 
\\
 \includegraphics[width=.5\columnwidth]{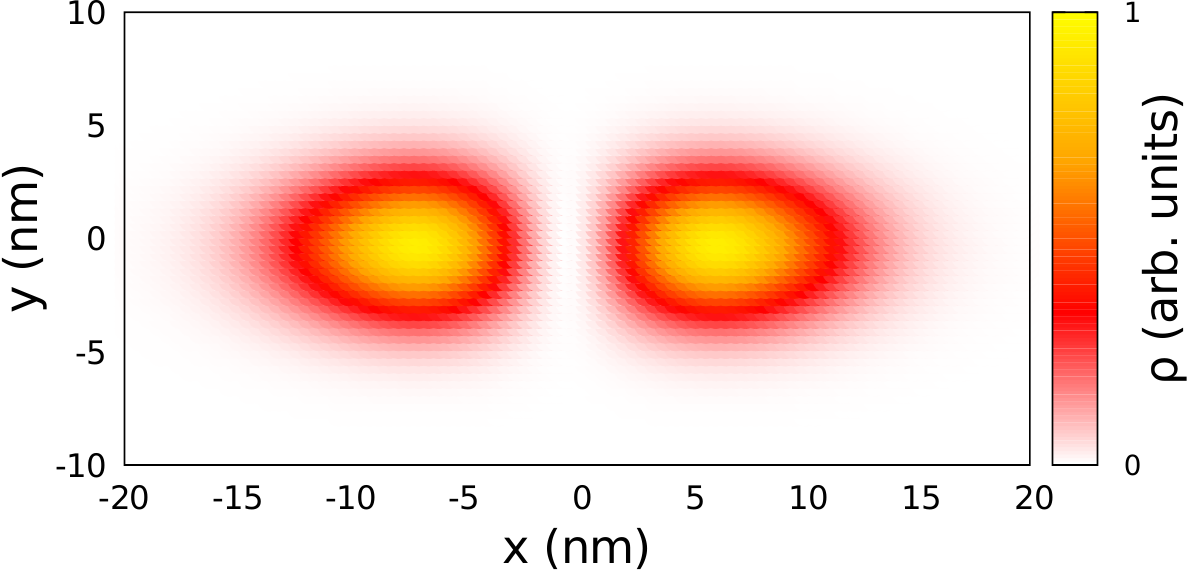} \put(-105,43){(e)} &
 \includegraphics[width=.5\columnwidth]{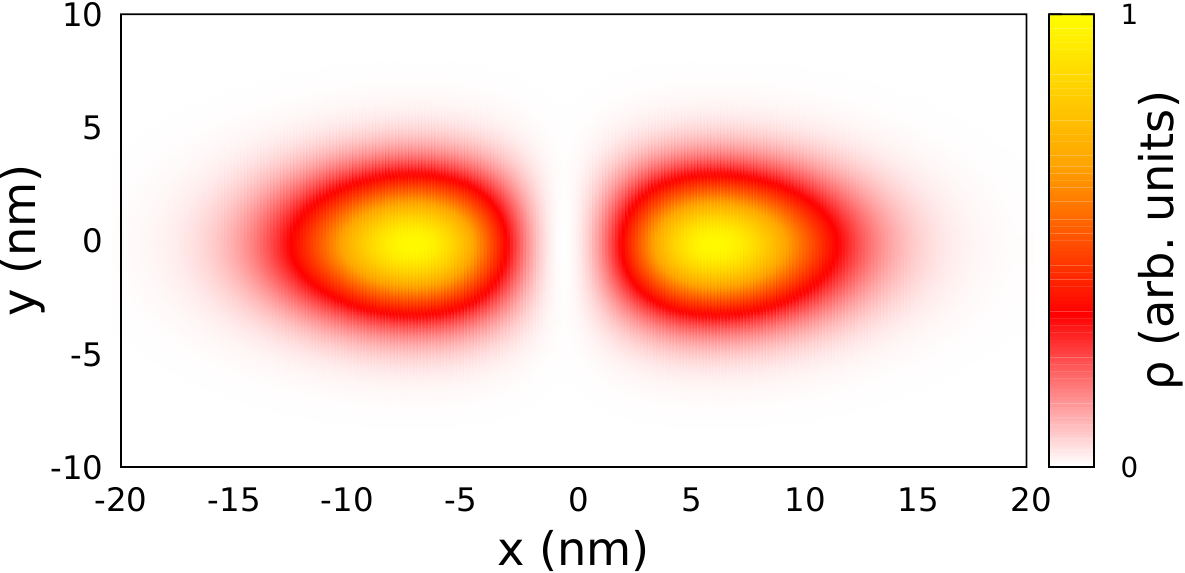}\put(-105,43){(f)}   

\end{tabular}

\caption{
Probability density for the single-electron  states
calculated with the tight-binding -- left column (a,c,e) and effective-mass approximation -- right column (b,d,f).
Panels (a,b) correspond to the single-electron ground state.
Second (c,d) and third row (e,f)  of plots correspond to the first and second excited single-electron states. 
}
 \label{szeff}
\end{figure}

 \begin{figure}
\begin{tabular}{ll}
 \includegraphics[width=.5\columnwidth]{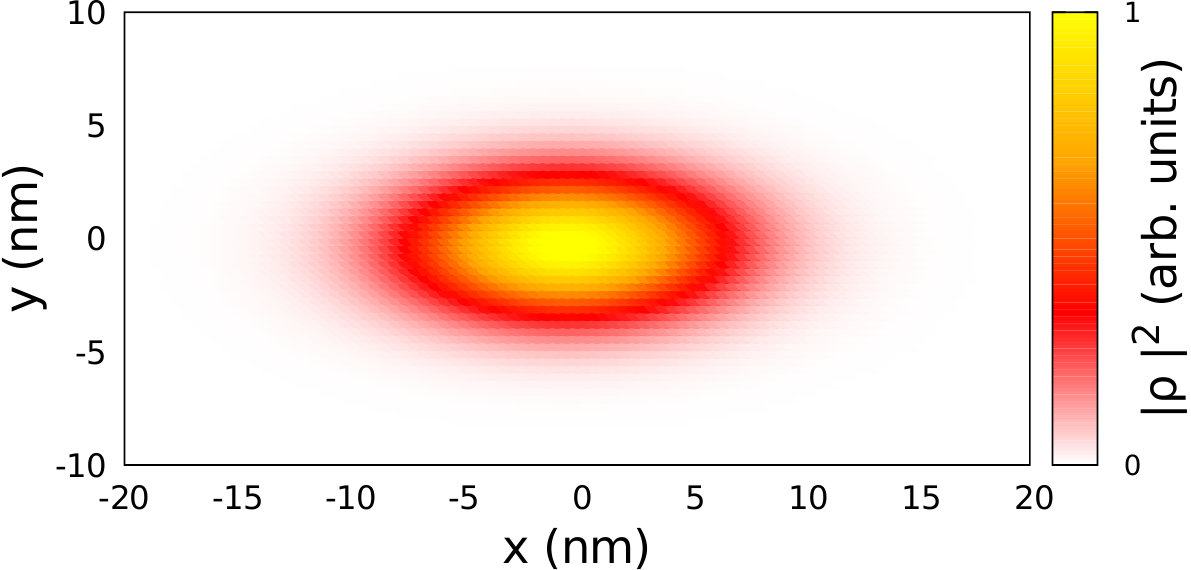}\put(-105,42){(a)} &
 \includegraphics[width=.5\columnwidth]{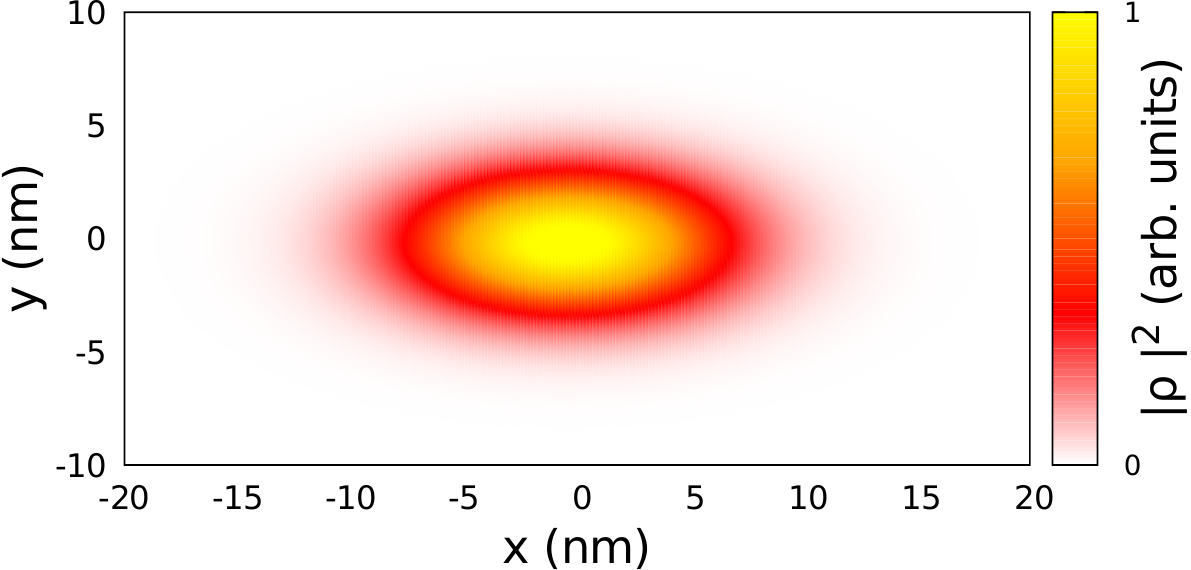}\put(-105,42){(b)} 
\\
\includegraphics[width=.5\columnwidth]{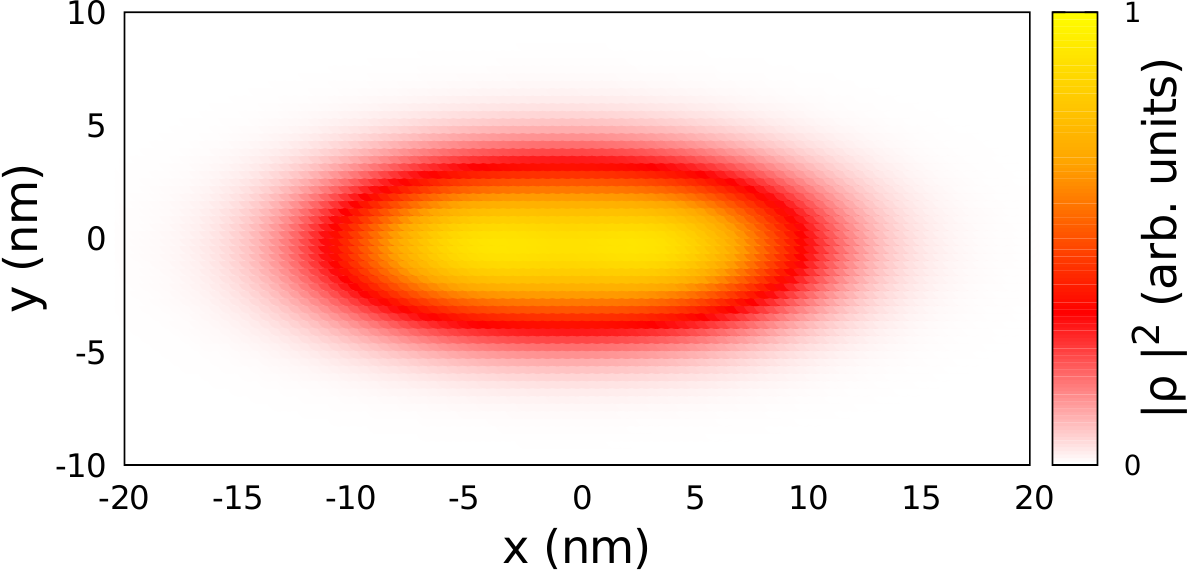}\put(-106,44){(c)} &
 \includegraphics[width=.5\columnwidth]{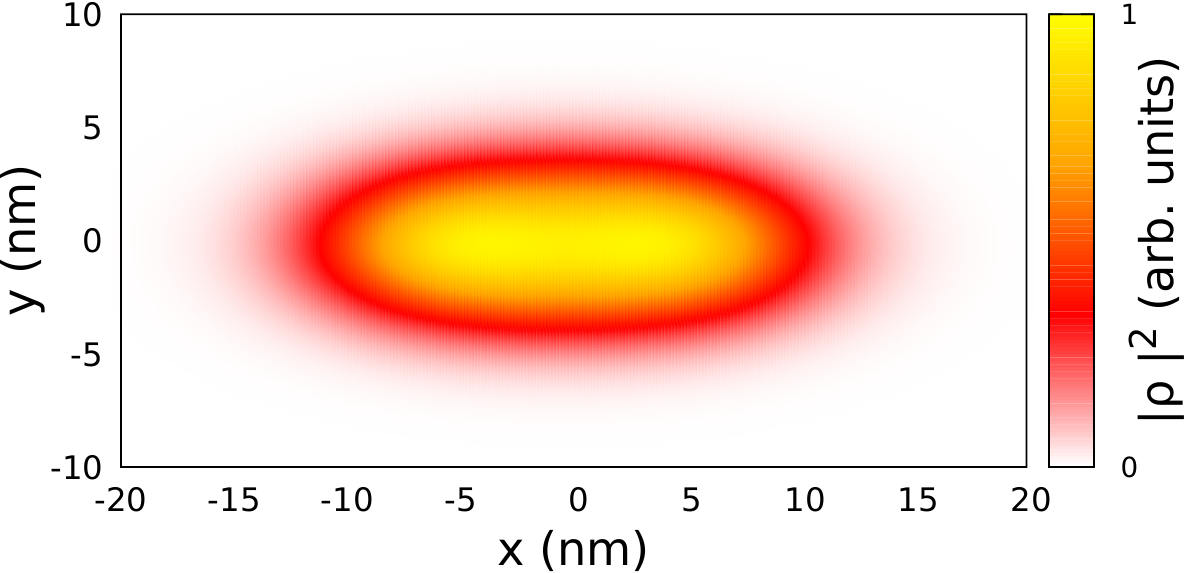}\put(-106,44){(d)} \\

\includegraphics[width=.5\columnwidth]{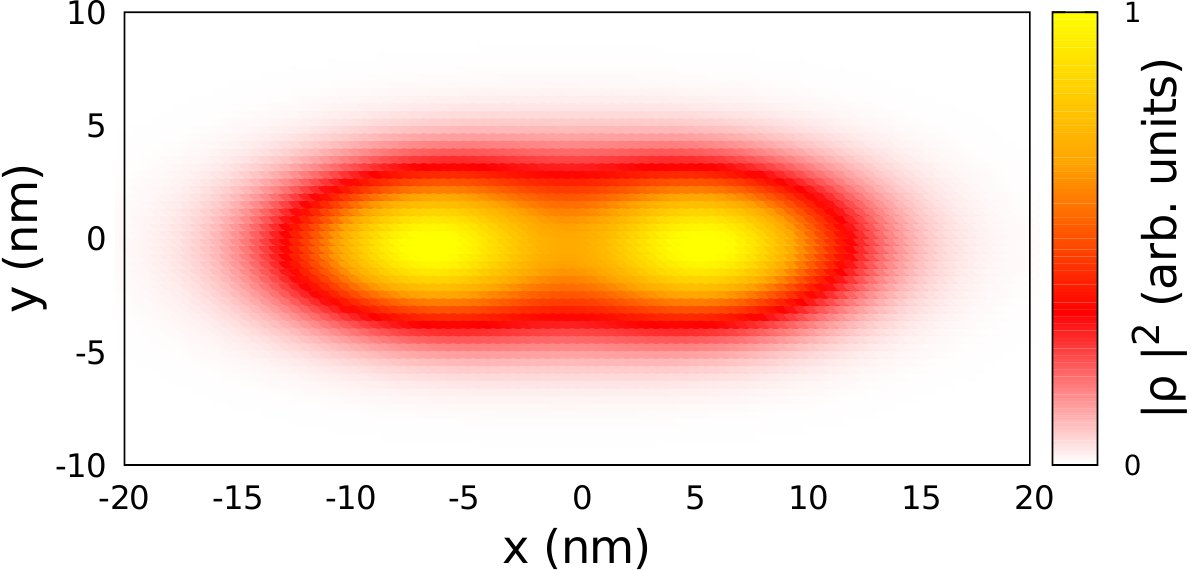}\put(-106,44){(e)} &
 \includegraphics[width=.5\columnwidth]{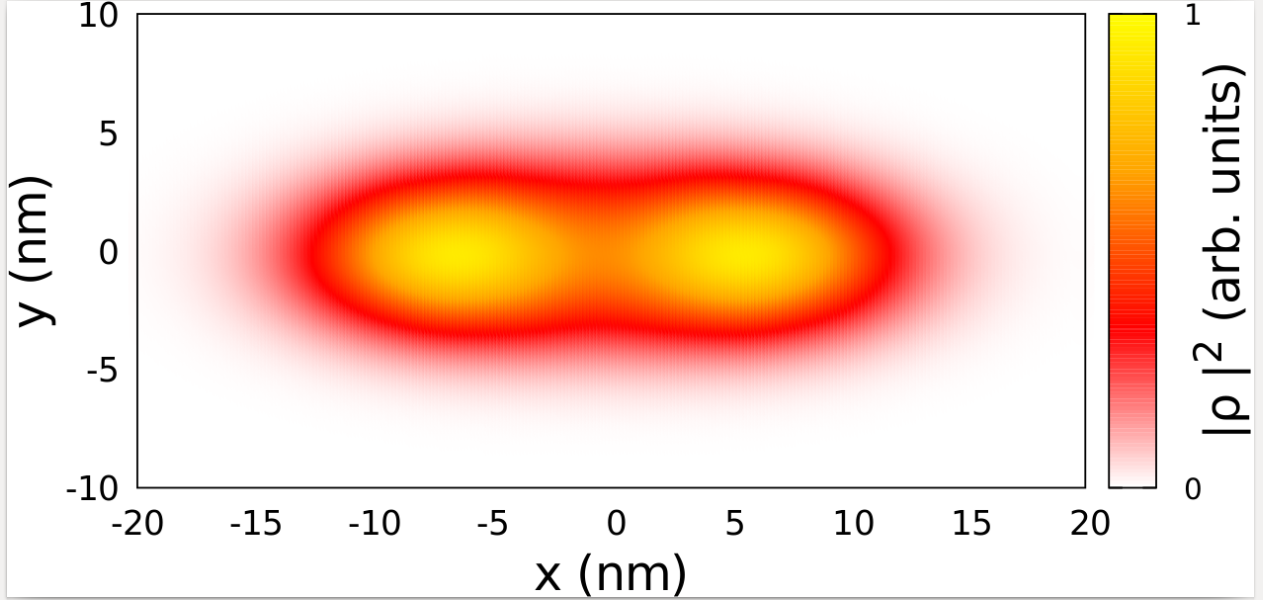}\put(-106,44){(f)} 
\\
 \includegraphics[width=.5\columnwidth]{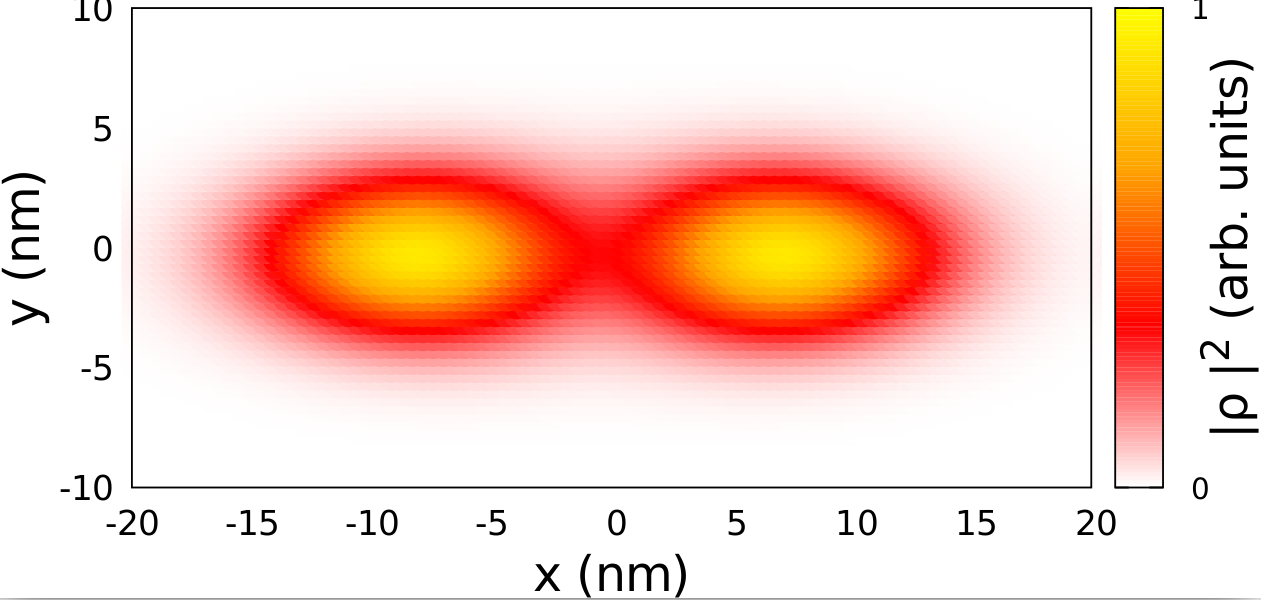}\put(-106,44){(g)} &
\includegraphics[width=.5\columnwidth]{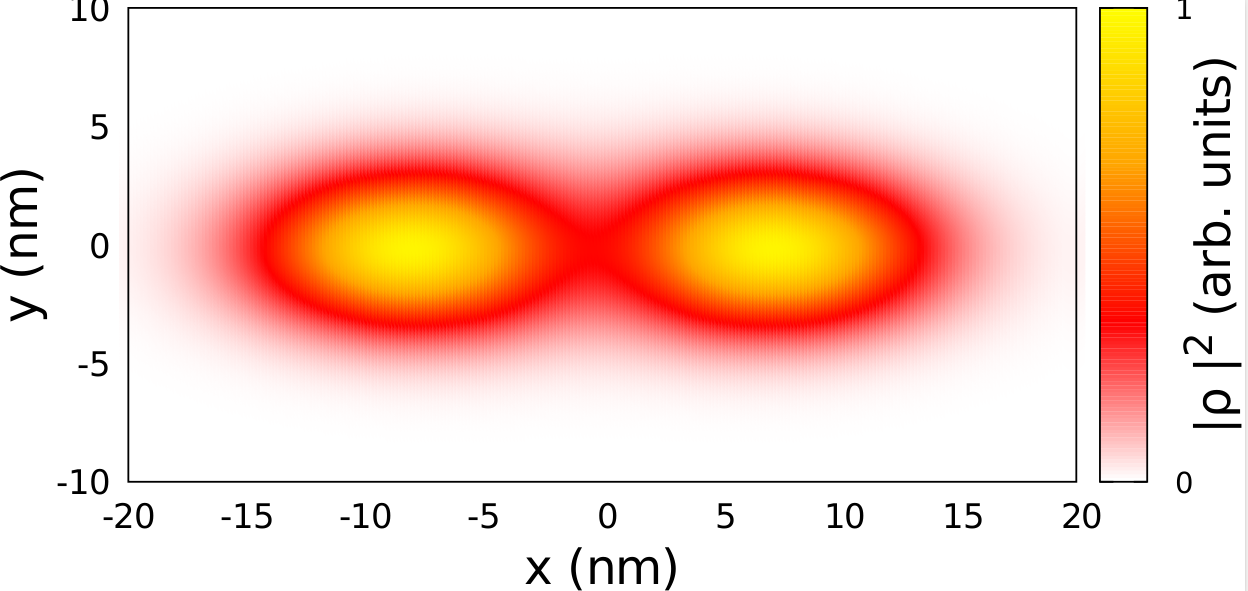}\put(-106,44){(h)}  \\
\end{tabular}

\caption{
Probability density for two-electron ground-states
calculated with the tight-binding -- left column -- (a,c,e,g,i) and effective-mass approximation --right column (b,d,f,h,j).
Panels  correspond to decreasing dielectric constant from top to bottom
$\epsilon=\infty$ (a,b), $\epsilon=18.2$  (c,d), $\epsilon=9.1$  (e,f), and $\epsilon=4.55$  (g,h).
In bulk of this work we apply $\epsilon=9.1$.}
 \label{szeff2}
\end{figure}

\begin{table}
\begin{tabular}{llll} 
$n$ & S/T & $\Delta E_{TB}$ (meV) & $\Delta E_{EM}$ (meV) \\ 
\hline
1 & S & 0 & 0  \\
2 & T & 1.228 & 1.175  \\
3 & T & 6.737 & 6.737  \\    
4 & S & 10.024 &10.002\\    
5 & S & 10.043 & 10.011  \\    
6 & S & 10.055 & 10.034  \\    
7 & T & 11.251 & 11.192  \\
8 & T & 11.312 & 11.249  \\   
9 & S & 12.673 & 12.656  \\  
10& T & 16.760 & 16.750 \\ 
\end{tabular}
\caption{The energy spacings in meV from the ground-state as calculated with the tight-binding
method ($\Delta E_{TB}$) and with the effective-mass approximation ($\Delta E_{EM}$) for 
potential of Eq. \ref{pote} at $B=0$. The second column indicates the spin-singlet states by S
and spin-triplet states by T.
} \label{tab3}
\end{table}

 \begin{figure}
\begin{tabular}{ll}

 \includegraphics[width=.52\columnwidth]{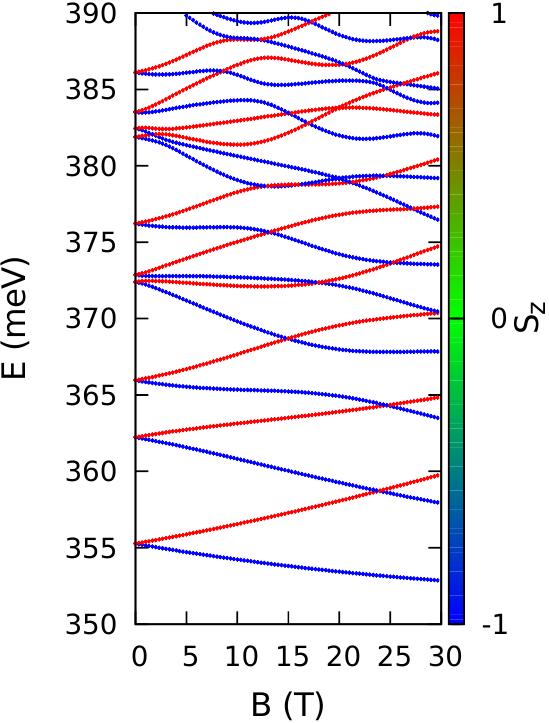} \put(-95,50){(a)} &
  \includegraphics[width=.5\columnwidth]{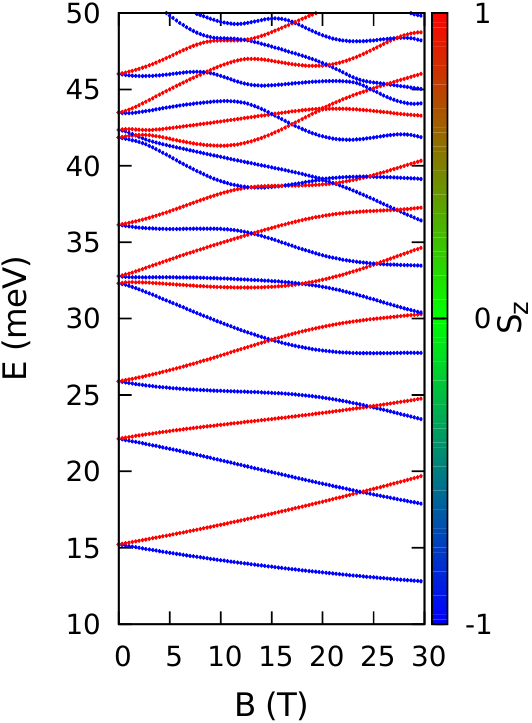}\put(-95,50){(c)}  \\
 \includegraphics[width=.52\columnwidth]{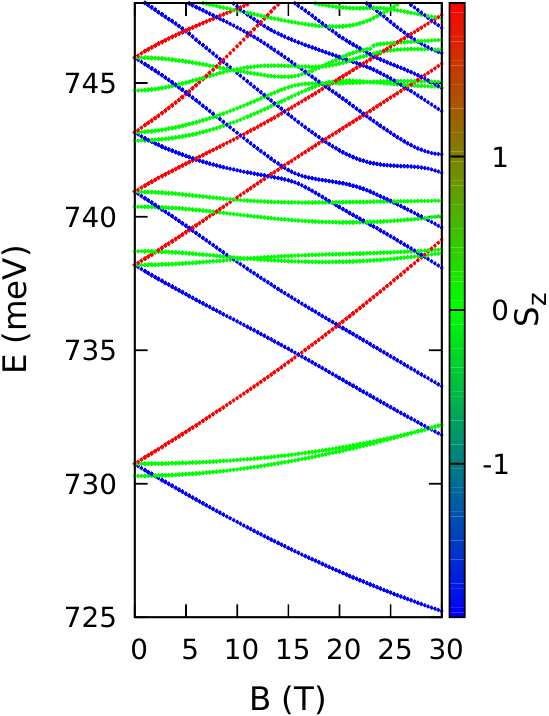}\put(-95,30){(b)} &
 \includegraphics[width=.5\columnwidth]{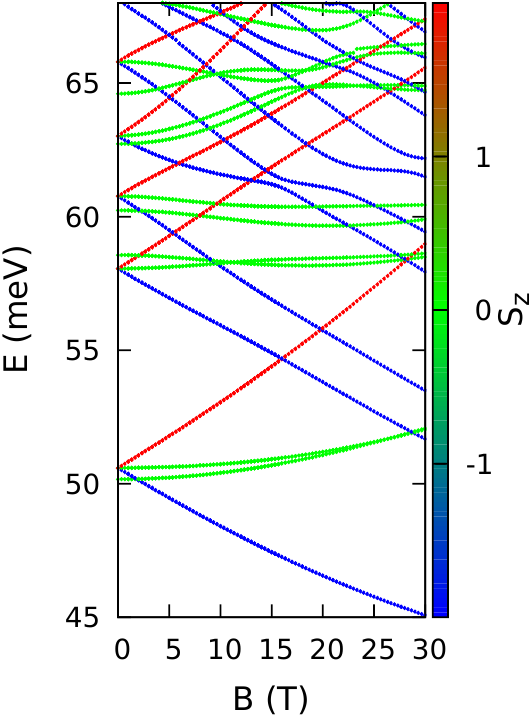}\put(-95,30){(d)} 
\\
\end{tabular}
\caption{Same as Fig. \ref{jedenidwae} but for a Gaussian perturbation to the parabolic potential
given by Eq. (\ref{potz}). }
 \label{zab}
\end{figure}

 \begin{figure}
\begin{tabular}{ll}
 \includegraphics[width=.5\columnwidth]{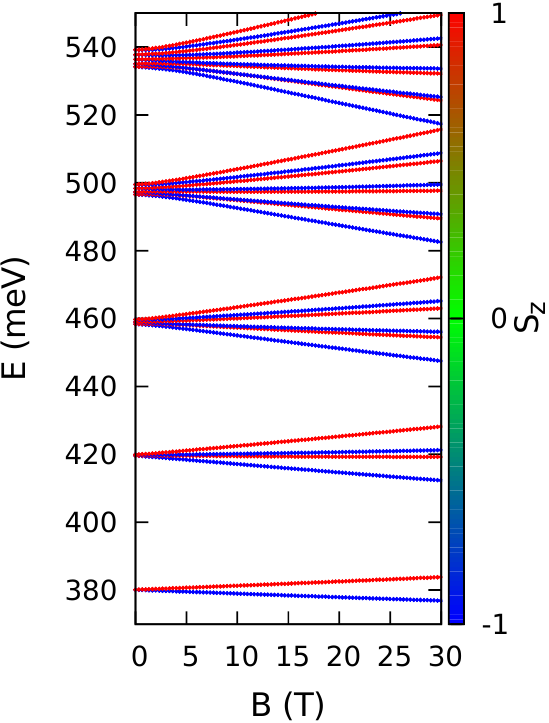}\put(-90,42){(a)} &
 \includegraphics[width=.5\columnwidth]{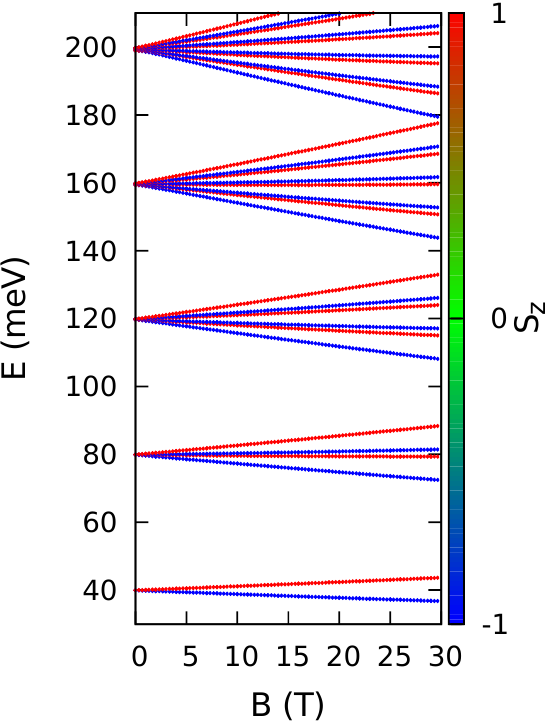}\put(-90,42){(b)} 
\end{tabular}

\caption{
Tight-binding (a) and effective mass (b) single-electron spectrum for $\hbar\omega=40$ meV.}
 \label{40}
\end{figure}

 \begin{figure}
\begin{tabular}{ll}
 \includegraphics[width=.5\columnwidth]{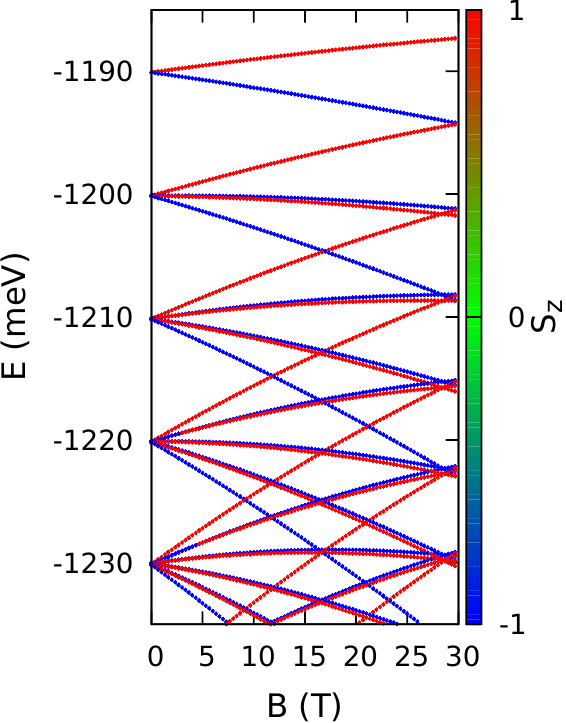}\put(-90,132){(a)} &
 \includegraphics[width=.47\columnwidth]{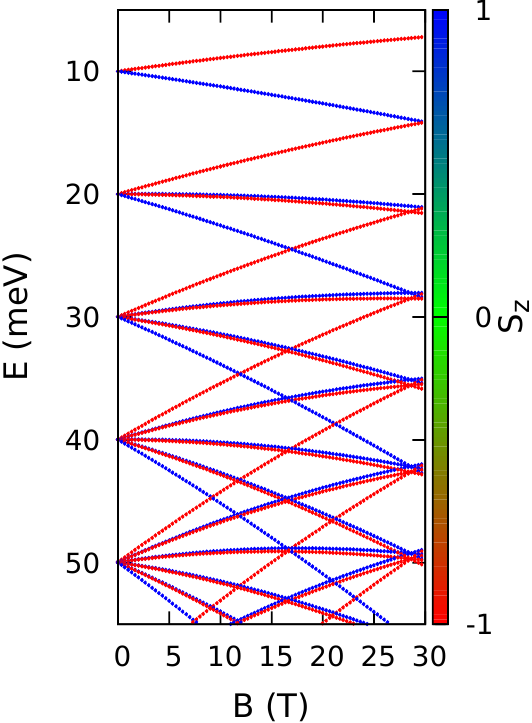}\put(-90,132){(b)} 
\end{tabular}

\caption{(a) The tight-binding energy spectrum for the valence-band confined states 
for inverted harmonic oscillator confinement with $\hbar\omega=10$ meV.
(b) A single-particle spectrum in the effective-mass approximation (note the inverted scales for 
the energy and the spin). The results were obtained for hole effective masses obtained by a fit
to the harmonic oscillator spectrum with $m_x=0.18972m_0$ and $m_y=1.15071 m_0$.}
 \label{hole}
\end{figure}

\subsubsection{Spectra in external magnetic-field and charge densities }

The dependence of the single-electron energy levels in the external magnetic field is given in Fig. \ref{jedenidwae}(a,c) with the same scale applied for the vertical and horizontal axes for the tight-binding [Fig. \ref{jedenidwae}(a)]  and
continuum approach [Fig. \ref{jedenidwae}(c)]. 
The probability density calculated with the tight-binding approach for the spin-down states:
the ground state, the  first  and the second excited states are displayed in Fig. \ref{szeff}(a,c) and (e), respectively.  For comparison the probability densities calculated with the effective-mass approximation are given in 
Fig. \ref{szeff}(b,d,f). The probability densities for these low-energy states are elongated along the armchair ($x$) direction and more strongly localized in the zigzag direction ($y$) and are very similar 
for the tight-binding Hamiltonian (left panels of Fig. \ref{szeff}) and in the single-band approximation (right panels of Fig. \ref{szeff}). 

The energy spectra for the electron pair are plotted in Fig. \ref{jedenidwae}(b,d). 
 A more detailed comparison is given in Table \ref{tab3} where
the spacing from the ground state is given along with the information whether the state
is spin-singlet or spin-triplet. The singlet-triplet order of the tight-binding energy levels is  reproduced
by the effective-mass approximation and the energy spacings as calculated
by the continuum method agree with a precision of about 0.06 meV to the tight-binding results. 
In Fig. \ref{szeff2} we compare the results for the ground-state two-electron density
as obtained with the tight-binding (left panels) and the continuum approach (right panels). 
Due to the large mass in the zigzag direction the ground states are localized near the $y=0$ line,
to a quasi-one dimensional channel that promotes separation of single-electron charges.
The panels from top to bottom of Fig. \ref{szeff2} correspond to decreasing dielectric constant.
The results in Fig. \ref{szeff2}(e,f) correspond to $\epsilon=9.1$ used in the bulk of this paper.
As the screening of the interaction is reduced the charge density undergoes 
 Wigner crystallization, with formation of the single-electron islands separated along the armchair direction ($x$ axis). Results of both the tight-binding  and the effective-mass approaches  are again very similar.

\subsubsection{Non-parabolic confinement}
In order to verify the effective-mass approximation for non-parabolic spectrum we introduced a perturbation to the potential given by Eq. (\ref{pote}) introducing a repulsive Gaussian centered at the $y$ axis, i.e. 
for potential 
\begin{equation} V_p(x,y)= V(x,y) + V_z\exp(-\frac{(y-y_s)^2+x^2}{R^2}),\label{potz} \end{equation}
where $V(x,y)$ is given by Eq. (\ref{pote}), $V_z=30$ meV and $y_s=R=5$ nm. 
Comparison of the spectra is given in Fig. \ref{zab} for the single electron [Fig. \ref{zab}(a,c)] and 
for the electron pair [Fig. \ref{zab}(b,d)]. 
For the single-electron we notice that the perturbation that lowers the symmetry of the potential results in reduction of the degeneracy found for $B=0$. The energy levels enter into avoided crossings that replace the crossings of the high-symmetry case of Fig. \ref{jedenidwae}(a,b). 
  For the electron pair we find that the perturbation leads to a reduced  triplet-singlet energy difference in the ground-state for $B=0$. The potential defect introduced above the potential minimum enhances the separation of the charges for the electron pair which  
  Wigner localization in 
 quasi one-dimensional systems results in the singlet-triplet degeneracy \cite{1dszafran} for fully separated single-electron islands. 
Overall agreement between the results of the effective-mass and tight-binding models is very good also for nonparabolic external potential.

\subsubsection{Strong confinement}

We find that the good quality of the effective-mass approximation holds  for $\hbar\omega \leq 10$ meV.
Deviations are found for extremely strong confinement. 
In the limit of strong confinement the effective-mass approximation 
can be expected fail since (i) the number of atoms in the confinement region becomes
limited and (ii) the stronger confinement induces larger contribution from Bloch states wave vectors far from the 
conduction band minimum where the non-parabolicity gets stronger. The effect (ii) is also induced by higher energy of the confined states. 
We calculated the electron energy spectrum for the harmonic oscillator energy increased
from $\hbar \omega=10$ meV to $\hbar \omega=40$ meV. 
In the tight-binding spectrum [Fig. \ref{40}(a)] a lifted degeneracy of higher energy levels at $B=0$ is evident
already at the scale of the figure. The splitting width of $n=2,3,4$ and 5 shells at $B=0$ is 0.3, 1.4, 2.9 and 5.0 meV,
respectively. Moreover, a stronger non-linear contribution to energy level dependence on the magnetic field at $B=0$ is distinct for higher energy levels in the tight-binding model as compared to the results of the effective-mass model [Fig. \ref{40}(b)].  With the conclusion that the effective-mass model works poorly for $\hbar\omega=40$ meV we should keep in mind that the single-electron energy spacings in electrostatic quantum dots is usually of the order of a few meV at most \cite{atmost}.

\subsection{Confined states of the valence band}
Electrostatic quantum dots can store either electrons or holes depending on the sign of the confinement potential.
In order to obtain confinement for carriers of the valence band we inverted the potential given by Eq. (2),
adopting $V(x,y)=-\frac{1}{2} m^h_x\omega^2 x^2-\frac{1}{2}m^h_y\omega^2 y^2$ for the external potential. In the middle of the phoshorene flake a maximum of the potential is produced that can store holes of the valence band. 
We performed fit of the effective masses in the way explained above for the electrons. The best fit to the harmonic oscillator spectrum is obtained for the hole masses  $m^h_x=0.18972m_0$, $m^h_y=1.15071 m_0$.
The comparison of the energy levels at $B=0$ is given in Fig. \ref{tab3}. 
For $n=1$ the deviation from the harmonic-oscillator levels given in the last column of Table \ref{tab3} is 2.8 times larger for the hole than for the electron (cf. Tables \ref{tab2} and \ref{tab3}). For harmonic oscillator shells  $n=2$, $3$ and $4$ 
the ratio of the deviations for the hole and for the electron is 4.16, 2.6 and 1.04, respectively.
The poorer performance of the effective-mass method for the hole is due to the stronger non-parabolicity of the valence band, particularly in the zigzag direction \cite{tibikast1}.
The hole masses derived from the band structure in the literature are: $m^h_x/m_0=0.138$ \cite{nano},0.15 \cite{18,016}, 0.182\cite{kdotp6}, and  $m^h_y/m_0=1.12$ \cite{18}, 1.14 \cite{kdotp6}, 4.92 \cite{016}, 5.917 \cite{nano}.
The strain strongly modifies the value of the hole masses \cite{nano}.
The extraction of the effective masses in the tight-binding approach for the flat valence band strongly depends on the way the fit of the parameters is produced, with narrow or wide range of the wave vectors near the $\Gamma$ point \cite{tbp}. In the present work the localization of the states in the $k$ vector space is determined by the localization of wave functions. 

On the scale of the Fig. \ref{hole} the results for the tight-binding and the effective-mass models are very similar.
The results for the latter model [Fig. \ref{hole}(b)] were obtained for confinement potential of Eq. (2) without its inversion. 
Instead, for comparison we inverted the scales for the energy and the spin in Fig. \ref{hole}(b).

\begin{table}
\begin{tabular}{l|lllll|l} 
$n\hbar \omega$ & $\Delta E_{TB}$  & &&&&$\sqrt{\sum_i (E^{ex}_i-E^{tb}_i)^2}$  \\
\hline
 10 & 10.009 & 10.093 &&&& 0.0934 \\  20 & 20.009 & 20.059 & 20.137 &&&0.1374\\ 30&  29.992 & 30.011 & 30.078 & 30.139&&0.1599  \\  40 & 39.929 &39.930 & 40.034 &40.059 & 40.117& 0.1682 \\
\end{tabular}
\caption{Same as Table \ref{tab2} only for the valence band with the fitted hole effective masses $m_x^h=0.18972m_0$, $m_y^h=1.15071 m_0$. 
} \label{tab3}
\end{table}

\section{Summary and Conclusions}
We have determined the energy spectra of carriers confined in a parabolic external potential using the effective tight-binding Hamiltonian for phosphorene, including the effects of the external magnetic field and the electron-electron interaction. We determined the effective masses for which the lowest energy levels as calculated with the tight binding approximately reproduce the exact quantum harmonic oscillator spectrum. The fitted masses were then used for 
a simple single-band effective-mass model for conduction band electrons and the valence band holes. 
The Wigner crystallization of the two-electron ground-state is found at $B=0$ already for a small quantum dot size. The effective mass model correctly reproduces the sequence of the singlet and triplet energy levels and the interaction effects.
The present demonstration that the states confined by electrostatic potentials in phosphorene can be with a good approximation described within the simple single-band effective-mass model  opens perspectives for simplified treatment of electrostatic quantum dots in monolayer black phosphorus.

\end{document}